\documentclass[aip,amsmath,amssymb,reprint]{revtex4-1}
\usepackage{graphicx}
\usepackage{dcolumn}
\usepackage{bm}
\usepackage[utf8]{inputenc}
\usepackage[T1]{fontenc}
\usepackage{mathptmx}
\usepackage{etoolbox}
\usepackage{booktabs,multirow}
\usepackage{makecell}
\usepackage{setspace}
\usepackage{diagbox}
\usepackage{setspace}
\usepackage{threeparttable}
\usepackage{color}
\makeatletter
\def\@email#1#2{%
	\endgroup
	\patchcmd{\titleblock@produce}
	{\frontmatter@RRAPformat}
	{\frontmatter@RRAPformat{\produce@RRAP{*#1\href{mailto:#2}{#2}}}\frontmatter@RRAPformat}
	{}{}
}%
\makeatother
\begin{document}
	\preprint{AIP/123-QED}
	
	\title{Separating micrometer-sized particles utilizing a dusty plasma ratchet}
	
	\author{Zhi-min Cai}
	\affiliation{ College of Physics Science and Technology, Hebei University, Baoding 071002, China}
	\author{Zong-bo Ma}
\affiliation{ College of Physics Science and Technology, Hebei University, Baoding 071002, China}
	\author{You-kai Zhao}
\affiliation{ College of Physics Science and Technology, Hebei University, Baoding 071002, China}
	\author{Fu-cheng Liu}
\affiliation{ College of Physics Science and Technology, Hebei University, Baoding 071002, China}
	\author{Ya-feng He}%
	\affiliation{ College of Physics Science and Technology, Hebei University, Baoding 071002, China}
	\affiliation{ Hebei Research Center of the Basic Discipline for Computational Physics, Hebei University, Baoding 071002, China}
	\email{heyf@hbu.edu.cn}
	
	\date{\today}
	
\begin{abstract}
Directional transport-dominated particle separation presents major challenges in many technological applications. The Feynman ratchet can convert the random perturbation into directional transport of particles, offering innovative separation schemes. Here, we propose the design of a dusty plasma ratchet system to accomplish the separation of micron-sized particles. The dust particles are charged and suspended at specific heights within the saw channel, depending on their sizes. Bi-dispersed dust particles can flow along the saw channel in opposite directions, resulting in a perfect purity of particle separation. We discuss the underlying mechanism of particle separation, wherein dust particles of different sizes are suspended at distinctive heights and experience electric ratchet potentials with opposite orientations, leading to their contrary flows. Our results demonstrate a feasible and highly efficient method for separating micron-sized particles.
\end{abstract}
	
\maketitle
	
\indent Particle separation has significant applications across various fields including physics, chemistry, biology and more. Numerous efforts have been devoted to developing effective separation methods based on different mechanisms, such as the sieve filter,\cite{Stogin} electrophoresis,\cite{Cui} acousticophoresis,\cite{Garg,Dai} and chromatography.\cite{Wei} Of particular interest is the Feynman ratchet, which has the remarkable ability to convert random motion into directional transport of particles.\cite{Hanggi,Reimann,Drexler,Mukhopadhyay} Ratchet-based particle separations have been achieved experimentally, such as in microfluidic ratchets,\cite{Matthias, Tang, Skaug,Roth, Kowalik}  and numerically based on Langevin\cite{Slanina,Reguera, Germs} or Monte Carlo\cite{Li} simulations. Compared with extensively studied neutral agents like the biological cells, if the particle is charged, especially those of micron-size, ratchet-based approaches would become versatile and practical in manipulating particle transport and separation by applying a ratchet electric potential, that is a challenging task.

\indent Dusty plasma is a system in which micron-sized dust particles are charged in a weakly ionized gas.\cite{Morfill, Shukla,Jorge} These dust particles can undergo transport, relying on the local plasma parameters which mainly determine the electric and ion drag forces, or through externally applied driving. Spontaneous long-rang transport between two dust sources has been observed.\cite{Edward} Tunable transport of dust particles has also been achieved by externally manipulating the potential profiles through the application of two consecutive harmonics excitation,\cite{Shinya} phase-controllable electrical electrodes,\cite{Jiang} or by utilizing a strong magnetic field.\cite{Land} Furthermore, particle separation of binary particle systems has been achieved under microgravity conditions, that was attributed to the process of uphill diffusion\cite{Killer} or a force disparity.\cite{Stefan} In a recent dusty plasma ratchet experiment,\cite{He} monodisperse dust particles were rectified into directional transport with a desirable direction by regulating the discharge conditions. We believe that the response of dust particles to the discharge conditions definitely depends on their sizes. Here, we report the experimental separation of bi-dispersed dust particles through the utilization of a straight ratchet in dusty plasma. The two sizes of dust particles, suspended at distinctive heights within the saw channel of the straight ratchet, can flow in opposite directions, resulting in a perfect separation purity of 100\%.

\begin{figure}[htbp]
	\begin{center}
		\includegraphics[width=8cm,height=6.5cm]{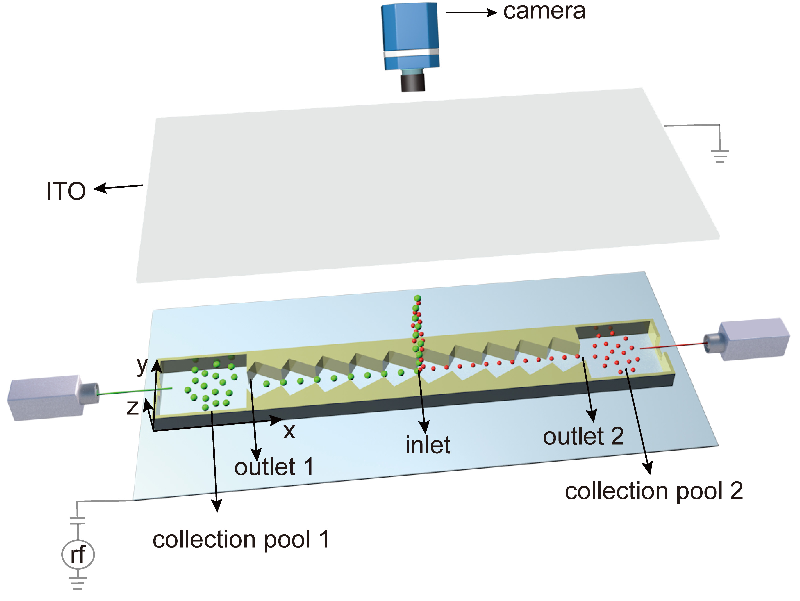}
		\caption{Sketch of the bi-dispersed particle separation process in a dusty plasma ratchet experiment. Bi-dispersed dust particles are injected at the inlet of the system. The small dust particles (red balls) and the large dust particles (green balls) flow in opposite directions towards the two outlets along the saw channel, and eventually into the two collection pools, resulting in the particle separation.}
	\end{center}
\end{figure}

\indent The experimental setup is depicted schematically in Fig.~1. A straight resin ratchet is positioned over the lower electrode connected to an $rf$ (13.56 MHz) power supply via a matching box. This ratchet structure comprises 26 sawteeth, with each sawtooth being 6 mm long, collectively extending over a length of 156 mm. The ratchet has a thickness of 9 mm, and the asymmetrical sawtooth possesses a depth of 3.5 mm. This configuration naturally forms a saw channel between the two side sawteeth of the straight ratchet, with the narrowest width of 6 mm. This saw channel is connected to two square collection pools embedded with slide glass at the bottom for gathering the separated dust particles. 

\indent Dust particles used are monodisperse polystyrene microspheres (from BaseLine ChromTech Research Center) with the relative standard deviation of the determination less than 4\%. To assess the separation effect of the dusty plasma ratchet, we select nine types of monodisperse dust particles with radii of 1, 2.5, 5, 7.5, 10, 11.5, 12, 13, and 14 $\mu$m, respectively. These two sizes are combined pairwise to demonstrate bi-dispersed particle separation.

\indent A capacitively coupled Argon plasma is generated between the upper and lower electrodes, which are separated by a distance of 60 mm. This plasma operates at typical gas pressures of $p$=10-30 Pa and discharge powers of $P$=5-30 W. To introduce binary dust particles into the system, we use a shaker initially containing a mixture of these particles, gradually dropping them into the saw channel through the middle inlet. These dust particles are confined to the center of the saw channel by the electric field of sheaths from the two sides sawteeth, and suspended at their respective balance heights, where the gravity and the upward electric field force balance, typically around $\sim$3-8 mm above the lower electrode, depending on their mass. The dust particles then perform specific flow along the saw channel until they ultimately reach and enter the collection pools. 

\indent To facilitate the observation of the separation process of the bi-dispersed dust particles, we use a red laser beam and a green laser beam, each with a power of 5 mW. The red laser beam is employed to illuminate and identify the small dust particles, while the green laser beam is used for the large dust particles. We utilize a camera (EOS 90D, 1920$\times$1080 pixels, 120 fps) from the upper transparent electrode to monitor and distinguish the movement of each size of dust particles during the particle separation process.

\begin{figure}[bp]
	\begin{center}\includegraphics[width=7.76cm,height=8cm]{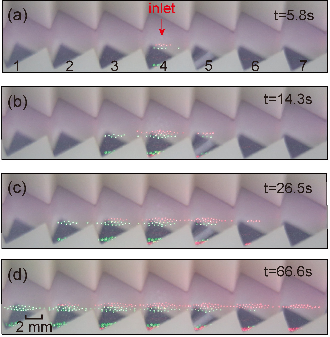}
		\caption{Particle separation experiments. (a) A small amount of a mixture of bi-dispersed dust particles with radii of $r_d$=7.5, 13 $\mu$m is injected into the inlet of the saw channel at the initial stage of the experiment. Small (large) dust particles are illuminated with a red (green) laser beam. (b)-(d) With the continual injection of the mixture, the bi-dispersed dust particles begin to flow along the saw channel, but in opposite directions, resulting in their separation. The visible particles at the bottom of each panel are the shadows of the laser illuminated particles suspended on the mirrored electrode plate. Gas pressure $p$=26 Pa, discharge power $P$=8 W. Multimedia view.}
	\end{center}
\end{figure}

\indent We first demonstrate the separation of a mixture of bi-dispersed dust particles with radii of $r_d$=7.5, 13 $\mu$m. Here, we gradually and carefully introduce this mixture into the saw channel through the middle inlet while simultaneously collecting the separated dust particles at the two collection pools, as depicted in Figs.~1 and 2 (Multimedia view). During the initial stage of the experiment, only a few dust particles are injected into the inlet of the saw channel, as indicated in Fig.~2(a). This limited number of particles prevents the formation of a significant collective effect, which is necessary to overcome the confining electric ratchet potential from the sawteeth and induce directional flow along the saw channel. As the number of dropped dust particles increases, the collective effect among charged dust particles becomes more pronounced. At this point, the confines of a single sawtooth region are no longer capable of accommodating all these charged dust particles. As a result, certain dust particles begin to move outside this sawtooth and enter the neighboring sawteeth, as shown in Fig.~2(b). This behavior is a consequence of the growing particle density and their interactions, leading to a more significant particle flow and redistribution throughout the saw channel. At this moment, the mixture of bi-dispersed dust particles starts to flow in the opposite direction, the small dust particles flow towards outlet 2, while the large dust particles transport towards outlet 1, suggesting a beginning of the particle separation.

\begin{figure}[htbp]
	\begin{center}\includegraphics[width=5.68cm,height=8cm]{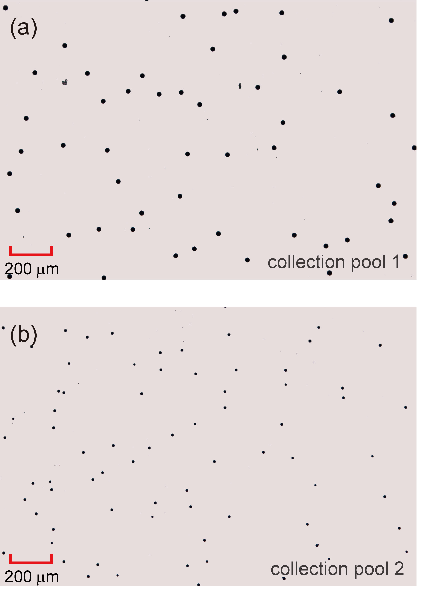}
		\caption{Micrograph showing the separation effect. (a) and (b) demonstrate the gathered large ($r_d$=13 $\mu$m) and small ($r_d$=7.5 $\mu$m) dust particles that fall on the two slide glasses located within the collection pools 1 and 2, respectively, indicating a separation purity of 100\%.}
	\end{center}
\end{figure}

\indent As we continuously drop the mixture of bi-dispersed dust particles into the inlet of the saw channel, the described process occurs repeatedly and successively. Consequently, the small dust particles consistently flow towards outlet 2, while the large dust particles persistently move towards outlet 1, as denoted in Figs.~2(c) and 2(d). It is conceivable that as long as we continuously drop the mixture of bi-dispersed dust particles into the inlet of the saw channel, we can achieve the final separation of the large dust particles at collection pool 1 and the small dust particles at collection pool 2, as demonstrated in Figs.~3(a) and 3(b). Our experiments clearly demonstrate the feasibility of using the dusty plasma ratchet as a method to separate micro particles effectively. 

\begin{figure}[htbp]
	\begin{center}
		\includegraphics[width=7.68cm,height=12cm]{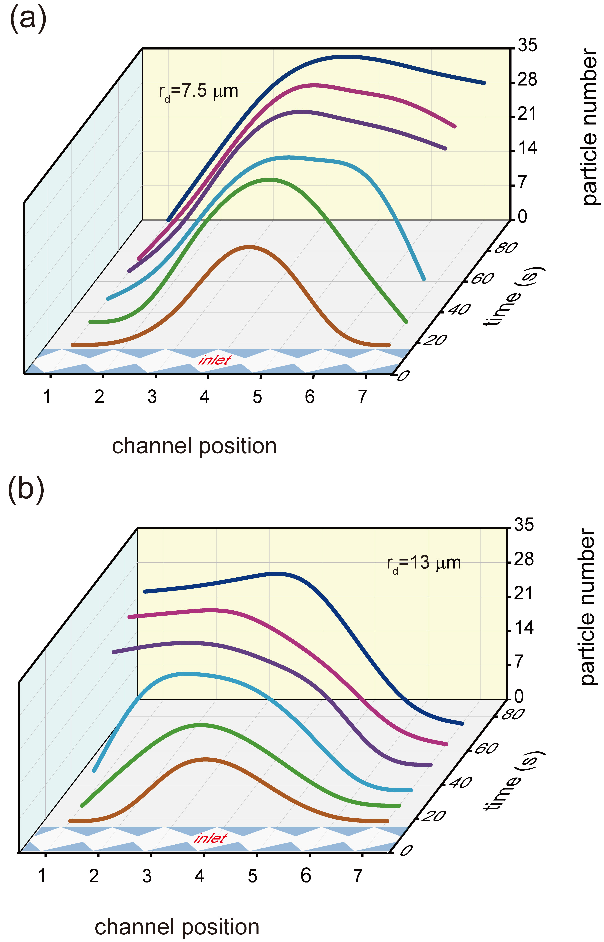}
		\caption{Distribution of particle number within the saw channel as a function of time during the continuous injection of the bi-dispersed dust particles.  As the experiment progresses, the small dust particles with a radius of $r_d$=7.5 $\mu$m flow towards outlet 2, (a), while the large dust particles with a radius of $r_d$=13 $\mu$m move towards outlet 1, (b).}
	\end{center}
\end{figure}

\indent Figure 4 illustrates the distribution of particle number within the saw channel as a function of time, providing crucial insights into the particle dynamics and the efficiency of the separation process. At $t$=18.6 s, when a small quantity of bi-dispersed dust particles enter the inlet, they are uniformly distributed near the inlet region. As additional bi-dispersed dust particles are continuously added into the inlet, the corresponding distributions for both the small and large dust particles extend towards both sides of the saw channel, as denoted in Figs. 4(a) and 4(b). This extension signifies the collective effect and increased particle density as more particles flow into the system. As the particle number continues to increase, the distribution of the small dust particles shows a lift at the right end, while the distribution of the large dust particles rises at the left end. These upward shifts in the curves represent the directional transport of the bi-dispersed dust particles in completely opposite directions.The shifting and evolution of the distribution curves over time clearly demonstrate the efficacy of the dusty plasma ratchet in achieving effective and directional particle separation. It's important to note that the peaks of the distributions of the bi-dispersed dust particles, which appear near the inlet, are a result of the continual injection of dust particles from this point. The continuous addition of particles at the inlet contributes to the steady flow and maintains the distributions throughout the process, leading to the particle separation observed in the experimental setup.

\begin{figure}[htbp]
	\begin{center}\includegraphics[width=7cm,height=6.4cm]{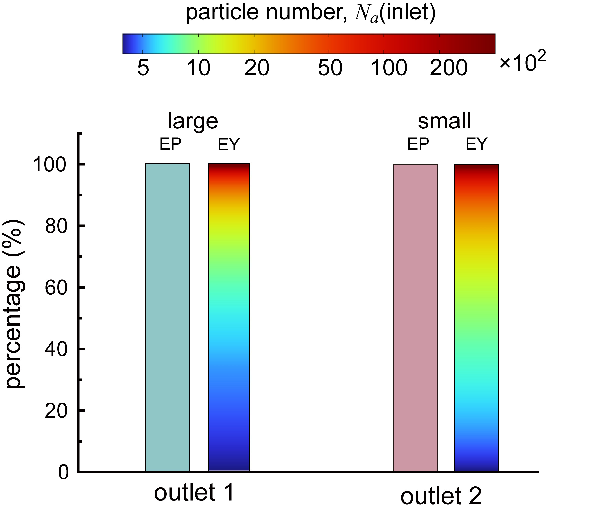}
		\caption{Separation purity and yield. At the outlet 1 (2), only large (small) dust particles are gathered with a purity of $EP$=100\%. The color-coded bars represent yields $EY$ which increase with the number $N_{a}{\rm(inlet)}$ of injected particles.}
	\end{center}
\end{figure}

\indent To evaluate the ability of our device to separate dust particles from a mixture, we calculate the purity $EP$, the proportion of particle type $a$ (either small or large) in outlet $i$ relative to the total number of particles in this outlet, and the yield $EY$, the number of particle type $a$ gathered in outlet $i$ over the total number of injected particles of the same type, defined as:
\begin{eqnarray}
	EP&=&\frac{N_a{\rm(outlet}_i)}{N_{total}{\rm(outlet}_i)}, \\
	EY&=&\frac{N_a{\rm(outlet}_i)}{N_{a}{\rm(inlet)}},
\end{eqnarray}
here, $N_a$ represents the number of particle $a$, $N_{total}$ the total particle number of the dropped mixture, $i$=1, 2 the outlet positions. The bar graphs $EP$ in Fig.~5 present the separation purity at the two outlets. At outlet 1, the collection is solely composed of large dust particles, indicating a purity of 100\%, as depicted in the micrograph presented in Fig.~3(a). Similarly, at outlet 2, a purity of 100\% is attained as well, with only small dust particles being gathered, as illustrated in the micrograph in Fig.~3(b). These outcomes highlight a perfect efficiency achieved in separating these dust particles. 

\indent The associated yields for the purified dust particles at the two outlets increase with the injected particle number, as depicted by the colored bars $EY$ in Fig.~5. This trend is attributed to the fact that a residual quantity of approximately $\sim$30 dust particles remains within each sawtooth, amounting to a total of approximately $N_{res}$$\sim$400 dust particles, after experimental operations. In Eq. (2), $N_{a}{\rm(outlet_i)}$ $=$ $N_{a}{\rm(inlet)}$ $-$ $N_{res}$, thus, increasing the injection number $N_{a}{\rm(inlet)}$ can significantly raise the yield. For instance, with an input of $N_{a}{\rm(inlet)}$$=$10,000 dust particles, a yield of 96\% can be achieved. In practical scenarios, where $N_{a}{\rm(inlet)}$ attains considerable magnitudes, a yield approaching 100\% becomes feasible.

\begin{table}
	\caption{\label{tab:table}Tested binary mixtures with particle radii $r_{d1}$ and $r_{d2}$.\\
			Y-separable.   N-inseparable in opposite directions}
\begin{ruledtabular}
			\begin{tabular}{c|ccccccccc}
				\diagbox { $r_{d1}$  ({\footnotesize $\mu$m})}{ $r_{d2}$ ({\footnotesize $\mu$m})}  & 1 & 2.5 & 5 & 7.5 & 10 & 11.5 & 12 & 13 & 14\\
				\hline
				1 & - & {\small N} & {\small Y}& {\small Y} & {\small Y} & {\small Y} & {\small Y} & {\small Y} & {\small Y}\\
				2.5 & {\small N}& - & {\small Y} & {\small Y} & {\small Y} & {\small Y} & {\small Y} & {\small Y} & {\small Y}\\
				5 & {\small Y}& {\small Y} & - & {\small Y} & {\small Y} & {\small Y} & {\small Y} & {\small Y} & {\small Y}\\
				7.5 & {\small Y}& {\small Y} & {\small Y}& -  & {\small Y} & {\small Y} & {\small Y} & {\small Y} & {\small Y}\\
				10  & {\small Y}& {\small Y} & {\small Y}& {\small Y} & -  & {\small N} & {\small N} & {\small Y}&{\small Y} \\
				11.5 & {\small Y}& {\small Y} & {\small Y}& {\small Y}  &{\small N} & - & {\small N} & {\small N}&{\small Y} \\
				12  & {\small Y}& {\small Y} & {\small Y}& {\small Y} &  {\small N} & {\small N}& - & {\small N}& {\small N} \\
				13  & {\small Y}& {\small Y} & {\small Y}& {\small Y} & {\small Y} & {\small N} & {\small N} & - & {\small N}\\
				14  & {\small Y}& {\small Y} & {\small Y}& {\small Y} & {\small Y} & {\small Y} &{\small N}  & {\small N} &- \\
			\end{tabular}
	\end{ruledtabular}
\end{table}

\indent We further examined the bi-dispersed particle separation using various binary mixtures with different particle radii, as detailed in Table 1. Generally, as long as the radius difference between the two dust particle sizes is larger than approximately 2 $\mu$m, the bi-dispersed dust particles can be directed to flow in opposite directions and effectively separated, ultimately gathering in their respective collection pools. However, when the radius difference between the two sizes of dust particles is very small, their suspended heights become so close that the positive ion wakes generated by the upper small particles start to influence the transport of the lower large particles. As a result, the bi-dispersed dust particles may not necessarily flow in opposite directions; instead, they could exhibit movement in the same direction but at different speeds. This implies a constraint on the effectiveness of the dusty plasma ratchet in achieving oppositely-flowing particle separation under these specific conditions.

\indent The proposed method has a stable separation efficiency for a wider range of particle radii. If one type of particles in the bi-dispersed mixture is sufficiently large, for instance, with $r_{d}$$\gtrsim$15 $\mu$m under the present experiment conditions, such that the electric field force can no longer support their weights, they would fall directly onto the lower electrode at the inlet. This situation is also applicable to other large and massive particles, such as metal and quartz. In this case, as long as the other type of particles in the bi-dispersed particle mixture can remain stably suspended and flow into the collection pool, the particle separation can still be effectively achieved.

\indent The observed directed flow of dust particles is unrelated to the used laser illumination, as the irradiation force exerted on the dust particles by the laser beam is completely negligible, owing to the laser power of 5 mW. To validate this, we conducted tests by interchanging the positions of the two lasers. It was observed that this manipulation had no impact on the directional transport and separation of dust particles. In addition, during the separation operation, it is necessary to gradually introduce the dust particles to establish directed flow, in order to avoid the rapid injection of a large number of dust particles surging towards the two ends of the saw channel.

\indent The particle separation demonstrated here results from the directional transport of the bi-dispersed mixture in opposite directions. The configuration of the straight ratchet creates a ratchet plasma sheath along the saw channel in the capacitively coupled discharge. Charged dust particles experience an electric ratchet potential within this non-electroneutral ratchet plasma sheath. The asymmetrical orientation of the ratchet potential determines the transport direction of dust particles. Due to the highly nonlinear feature of the sheath and the height-dependent transport property of dust particles, the small and large dust particles are suspended at different heights within the sheath and experience their respective ratchet potentials with opposite orientations, resulting in their opposite flow and the particle separation. Therefore, by controlling the discharge conditions to regulate the bi-dispersed mixture to be suspended at appropriate heights within the sheath, we can enable them to flow in opposite directions.

\indent In summary, we have realized the micron-sized particle separation using a designed dusty plasma ratchet. The height-dependent transport property of dust particles enables the bi-dispersed dust particles to flow in opposite directions along the straight ratchet. The designed dusty plasma ratchet demonstrated its capability and feasibility for separating micron-sized particles with a remarkable purity of 100\% and a higher yield. This research opens up promising possibilities for practical particle separation applications, with potential benefits in various fields such as microfluidics, nanotechnology, and particle manipulation. 

\indent  Project supported by the National Natural Science Foundation of China (Grant Nos. 11975089 and 12275064), the Interdisciplinary Research Program of Natural Science of Hebei University (Grant No. DXK202010), and the Hebei Natural Science Fund (Grant No. A2021201003), and the Scientific Research and Innovation Team project of Hebei University (Grant No. IT2023B03).
\\
\\
\noindent  \textbf{AUTHOR DECLARATIONS}\\
\textbf{Conflict of Interest}\\
\indent  The authors have no conflicts to disclose.\\

\noindent  \textbf{Author Contributions}\\
\noindent  \textbf{Zhi-min Cai:} Data curation (lead); Formal analysis (lead); Investigation (lead); Methodology (lead); Validation (equal); Visualization (equal); Writing – original draft (equal). \textbf{Zong-bo Ma:} Data curation (equal); Formal analysis (equal); Validation (equal); Writing – original draft (equal). \textbf{You-kai Zhao:} Data curation (equal); Formal analysis (equal); Validation (equal). \textbf{Fu-cheng Liu:} Data curation (equal); Formal analysis (equal); Supervision (equal); Validation (equal); Writing – review \& editing (equal). \textbf{Ya-feng He:} Data curation (equal); Funding acquisition (lead); Supervision (equal); Validation (equal); Writing – review \& editing (equal).\\

\noindent  	\textbf{DATA AVAILABILITY}\\
\indent  The data that support the findings of this study are available from the corresponding authors upon reasonable request.

\bibliography{aipsamp}

\end{document}